\documentclass[aps,pra,twocolumn,showpacs,floatfix]{revtex4-1}

\usepackage{bm,dcolumn,amsmath,graphicx}
\usepackage{epsfig}
\usepackage{nicefrac}
\def\cjp{Can. J. Phys.}

\def\epl{Europhys. Lett.}

\def\lnp{Lect. Notes Phys.}

\def\mnras{Mon. Not. R. Astron. Soc.}

\def\npa{Nucl. Phys. A}

\def\pra{Phys. Rev. A}

\def\prc{Phys. Rev. C}

\def\prl{Phys. Rev. Lett.}

\def\pscr{Phys. Scr.}

\def\sci{Science}

\newcommand{\paper}{paper}

\newcommand{\eref}[1]{(\ref{#1})}

\newcommand{\Tref}[1]{Table~\ref{#1}}
\newcommand{\Fig}[1]{Fig.~\ref{#1}}

\newcommand{\E}[1]{\ensuremath{\times 10^{#1}}}
\newcommand{\cm}{\ensuremath{\textrm{cm}^{-1}}}

\begin{document}

\title{Optical transitions in highly-charged californium ions with high sensitivity to variation of the fine-structure constant}

\author{J. C. Berengut}
\author{V. A. Dzuba}
\author{V. V. Flambaum}
\author{A. Ong}
\affiliation{School of Physics, University of New South Wales, Sydney, NSW 2052, Australia}
\date{April 3, 2012}

\pacs{06.30.Ft, 31.15.am, 32.30.Jc}

\begin{abstract}
We study electronic transitions in highly-charged Cf ions that are within the frequency range of optical lasers and have very high sensitivity to potential variations in the fine-structure constant, $\alpha$. The transitions are in the optical despite the large ionisation energies because they lie on the level-crossing of the $5\!f$ and $6p$ valence orbitals in the thallium isoelectronic sequence. Cf$^{16+}$ is a particularly rich ion, having several narrow lines with properties that minimize certain systematic effects. Cf$^{16+}$ has very large nuclear charge and large ionisation energy, resulting in the largest $\alpha$-sensitivity seen in atomic systems. The lines include positive and negative shifters.
\end{abstract}

\maketitle

\section{Introduction}
In this work we present calculations of transitions in highly-charged californium that could form the reference for an optical atomic clock with very strong sensitivity to variation of the fine-structure constant, $\alpha = e^2/\hbar c$. Our work is motivated by recent astronomical studies of quasar absorption spectra that indicate a spatial gradient in values of $\alpha$ across cosmological distances~\cite{webb11prl,king12mnras}. The results were taken using around 300 spectra covering most of the sky, observed at two telescopes: the Very Large Telescope in Chile~\cite{king12mnras} and Keck Telescope in Hawaii~\cite{webb99prl,murphy03mnras,murphy04lnp}. The telescopes independently agree on the direction and magnitude of the gradient (dipole), which is significant at $4.2\sigma$ for the combined sample of both telescopes.%~(see also~\cite{berengut12arxiv0}).

The cosmological dipole in $\alpha$ might be confirmed by terrestrial studies, since the solar system is moving with respect to the cosmic microwave background (the presumed frame for the $\alpha$-dipole), and therefore should be moving from a region of the Universe with smaller values of $\alpha$ to one with larger values~\cite{berengut12epl}. In particular, the expected rate of change in $\alpha$ today would be of order $\dot\alpha/\alpha \sim 10^{-18}\,\textrm{yr}^{-1}$. This is significantly smaller than the best current terrestrial limits, $\dot\alpha/\alpha = (-1.6\pm 2.3)\E{-17}\,\textrm{yr}^{-1}$, which comes from comparison of Al$^+$ and Hg$^+$ atomic clocks~\cite{rosenband08sci}. If measured at the same level of accuracy, the transitions proposed in this work would allow an improvement on this limit by a factor of 23. Because the transitions have narrow natural line widths and reduced systematics, the improvement could be even larger.

We parametrize the sensitivity of an atomic transition to potential variation in $\alpha$ by the quantity $q$ defined by
\begin{equation}
\label{eq:q_def}
q = \left. \frac{d\omega}{d x}\right|_{x=0}
\end{equation}
where $x = \alpha^2/\alpha_0^2 - 1$ is the fractional change in $\alpha^2$ from its current value $\alpha_0^2$, and $q$ and $\omega$ are measured in atomic units of energy. In the Al$^+$ and Hg$^+$ comparison, the Al$^+$ clock is an ``anchor'' (relatively insensitive to $\alpha$-variation) while the mercury clock has a strong sensitivity of $q = -52\,200\,\cm$~\cite{dzuba08pra0}. An approximate formula for the $q$ value of a single energy level ($E_n = -I_n$ where $I_n$ is the ionisation energy of the level) with effective principal quantum number $\nu$ and angular momentum $j$ is~\cite{dzuba99pra, berengut10prl}
\begin{equation}
\label{eq:q_approx}
q_n \approx - I_n \frac{(Z\alpha)^2}{\nu (j + 1/2)} \,,
\end{equation}
where $Z$ is the nuclear charge. The transition will have a sensitivity to $\alpha$-variation that is the difference between the $q$ values of the levels involved. Therefore the best transitions will maximise the difference of $\nu$ and $j$ between the levels and will come from heavy ions.

Equation~\ref{eq:q_approx} shows that transitions in highly charged ions (HCIs) can have much larger $q$ values since they have much larger ionisation energies. Unfortunately they generally also have much larger transition energies, putting them outside the range of optical lasers and making them unsuitable for use in high-precision clocks. However, due to configuration crossing, some HCIs can have optical transitions between levels with different principal quantum numbers, and these could become reference transitions for optical clocks with the highest $q$ values seen in atomic systems~\cite{berengut10prl,berengut11prl}.

\begin{table}[b]
\caption{Half-lives and symmetries of long-lived californium isotopes~\cite{audi03npa}.}
\label{tab:CfIso}
\begin{ruledtabular}
\begin{tabular}{ccc}
{Isotope} & $J^{\pi}$ &
{Half-life (yr)} \\
\hline
$^{248}$Cf & $0^{+}$ & 0.914\\
$^{249}$Cf & $9/2^{-}$ & 351\\
$^{250}$Cf & $0^{+}$ & 13.08\\ 
$^{251}$Cf & $1/2^{+}$ & 900\\
$^{252}$Cf & $0^{+}$ & 2.645
\end{tabular}
\end{ruledtabular}
\end{table}

In this \paper\ we present calculations for the $5\!f$ -- $6p_{1/2}$ crossing that occurs in the thallium and lead isoelectronic sequences (with one valence electron and two valence electrons, respectively). The crossing occurs at $Z = 98$ --- californium --- which also happens to be one of the last relatively stable ions in the periodic table, having isotopes with half-lives of several hundred years. The five isotopes shown in \Tref{tab:CfIso} allow for choice in experiments: isotopes with an odd number of nucleons have a greater half-life, but isotopes with an even number of nucleons exclude the possibility of hyperfine structure. We find that Cf$^{16+}$ has several transitions that may be suitable for studies of $\alpha$-variation, including those with the largest $q$ values yet found.

\section{Method}
\begin{figure}[tb]
\caption{Dirac-Fock energies of the $6p_{1/2}$ (diamonds, dashed line), $6p_{3/2}$ (crosses, dot-dashed line), and $5\!f$ (circles, solid line) levels in the thallium isoelectronic sequence with increasing nuclear charge. The inset shows an enlarged view of the crossing region.\label{fig:Tlcrossing}}
\includegraphics[width=0.47\textwidth]{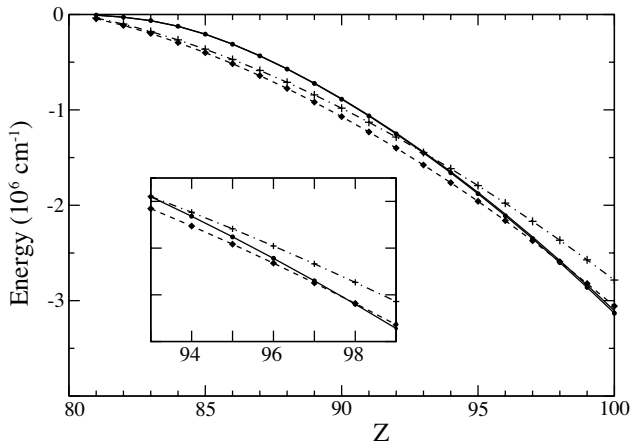}
\end{figure}

To find the $6p$ -- $5\!f$ crossing, we start with neutral thallium. In thallium, the $5\!f$ orbital energies lie above the $6p$ orbitals, whereas in the large $Z$ limit the $5\!f$ levels should be more tightly bound than $6p$ levels ($E_{5f} \approx E_{5p}$ for hydrogen-like ions). Therefore we expect a level crossing at some $Z > 81$, where an ion may have optical transitions between these two orbitals. \Fig{fig:Tlcrossing} shows the Dirac-Fock energies of the $6p_{1/2}$, $6p_{3/2}$, $5\!f_{5/2}$ and $5\!f_{7/2}$ orbitals as a function of nuclear charge $Z$. Due to the large fine-structure splitting of the $6p_{1/2}$ and $6p_{3/2}$ subshells, there are two possible crossing points we can explore here, one for $6p_{3/2}$ near $Z = 93$ and one for $6p_{1/2}$ near $Z = 98$. The crossing near $Z = 98$ is more attractive for studying $\alpha$-variation for two reasons. Firstly, the nuclear charge and ionization energy are larger. Second, since $\alpha$-sensitivity is due to relativistic effects that occur near the origin, the $6p_{1/2}$ orbital has larger $q$ than the $6p_{3/2}$ orbital since the former has a lower Dirac-spinor component of $s_{1/2}$ symmetry, which is large near the origin. This is seen in \eref{eq:q_approx} by the factor $1/(j+1/2)$: the difference in $q$ due to this factor is larger for $p_{1/2}$ and $f_{5/2}$. Because the $6p_{1/2}$ level is highly sensitive to $\alpha$-variation while the $5\!f$ levels are not, we expect a large $q$ value for a transition between these levels.

We have performed full-scale \textit{ab~initio} calculations for Cf$^{17+}$ (at the crossing point of \Fig{fig:Tlcrossing}) and the two-valence-electron equivalent, Cf$^{16+}$. We use the combined configuration interaction and many-body perturbation theory method (CI + MBPT), presented in detail in~\cite{dzuba96pra} (see also~\cite{berengut06pra}). We begin with Dirac-Fock for closed shells of Hg; this corresponds to V$^{N-1}$ for the single-valence-electron case, Cf$^{17+}$, and V$^{N-2}$ for the Cf$^{16+}$ ion. From the frozen-core potential we generate a set of around 40 $B$-splines in each wave up to $l = 6$. These form a ``complete'' set of virtual orbitals with which we calculate MBPT corrections, $\Sigma$, to second-order in the residual Coulomb interaction. For Cf$^{16+}$ we perform a CI calculation including all two-electron excitations to the virtual orbitals $16spdf$. The addition of $g$-wave orbitals to the CI were found to make little difference to energy levels and $q$ values. The $q$ values were obtained by varying $x$~(Eq.~\ref{eq:q_def}) in steps of $0.01$ and taking the gradient of transition frequency with respect to $x$. 

We have also calculated some important transition rates (reduced matrix elements and Einstein $A$-coefficients) using a relativistic formalism (see, e.g.~\cite{johnson07book}). Random phase approximation corrections to the matrix elements were not included in this work since the uncertainty in the rates is dominated by uncertainty in the transition energies, which have not been measured.

\section{Results and Discussion}
Our calculated energy levels and $q$ values for Cf$^{17+}$ are presented in \Tref{tab:Cf17+}. For the purposes of measuring $\alpha$-variation using atomic clocks, the most interesting transition is from the $5\!f_{5/2}$ ground state to $6p_{1/2}$, with an energy interval of $\omega = 17889\,\cm$. $q$ for this transition is over 8 times larger than the Hg$^+$ clock transition used in~\cite{rosenband08sci}.

\begin{table}[tb]
\caption{Low-lying levels of Cf$^{17+}$ (all have odd parity). Energy calculations are presented relative to the $5\!f_{5/2}$ ground-state using only Dirac-Fock (DF) and including MBPT (DF + $\Sigma$). The $q$ values were calculated using DF + $\Sigma$. }
\label{tab:Cf17+}
\begin{ruledtabular}
\begin{tabular}{lcrrr}
\multicolumn{1}{c}{Configuration} & $J$ &
\multicolumn{2}{c}{Energy~(\cm)} & 
\multicolumn{1}{c}{$q$~(\cm)} \\
&& DF & DF + $\Sigma$ &\\
\hline
$5\!f$ & 5/2 & 0     & 0     & 0 \\
$6p$   & 1/2 & 8447  & 17889 & -449750 \\
$5\!f$ & 7/2 & 20447 & 21755 & 17900 \\
$6p$   & 3/2 & 233514 & 241970 & -115650
\end{tabular}
\end{ruledtabular}
\end{table}

In Cf$^{16+}$, presented in \Tref{tab:Cf16+}, there are more states in optical range, arising from the greater number of angular momentum combinations available. Note that the levels marked A, A* in \Tref{tab:Cf16+} are heavily mixed in the CI calculation, resulting in a dominant contribution from $5\!f^2$ (63\% and 51\% respectively) in both, while in the CI + $\Sigma$ calculation the first state is 96\% $5\!f^2$ and the second state is 63\% $5\!f\,6p$. For simplicity we have simply labelled the level with the largest $5\!f\,6p$ contribution as the $5\!f\,6p$ state.

\begin{table}[tb]
\caption{Calculated energy levels, $g$-factors, and $q$ values for low-lying levels of Cf$^{16+}$ (all have even parity) relative to the ground state $5\!f\,6p\ (J=3)$. Energies are calculated using only configuration interaction (CI) and including many-body perturbation theory effects (CI + $\Sigma$). The ID column just provides a convenient label reference in the text.}
\label{tab:Cf16+}
\begin{ruledtabular}
\begin{tabular}{lccrrrr}
\multicolumn{1}{c}{Config.} & $J$ & ID & \multicolumn{1}{c}{$g$} &
\multicolumn{2}{c}{Energy~(\cm)} & 
\multicolumn{1}{c}{$q$~(\cm)} \\
&&&& CI & CI + $\Sigma$ &\\
\hline
$5\!f\,6p$ & 3 & G1 & 0.8299  & 0      & 0    & 0 \\
$6p^{2}$   & 0 & G2 &         & -7429  & 5267 & -370928 \\
$5\!f\,6p$ & 2 &  & 0.8482  & 7313   & 6104 & 106124 \\
$5\!f^2$   & 4 & A & 0.8535  & 28746 & 9711 & 414876 \\
$5\!f\,6p$ & 4 & A* & 1.0481 & 21415 & 24481 & 162126 \\
$5\!f^2$   & 2 &  & 0.7532  & 38674  &  24483 & 354444 \\
$5\!f\,6p$ & 3 &  & 1.1776  & 23979  & 25025 & 59395 \\
$5\!f^2$   & 5 &  & 1.0333  & 43097  & 29588 & 451455 \\
$5\!f^2$   & 3 &  & 1.0954  & 50953  & 37467 & 393755 \\
$5\!f^2$   & 4 &  & 1.1197  & 53229  & 42122 & 319216 \\
$5\!f^2$   & 6 &  & 1.1371  & 57220  & 44107 & 459347 \\
$5\!f^2$   & 0 & B &         & 68192  & 51425 & 380986 \\
$5\!f^2$   & 2 &  & 1.1672  & 67267  & 51471 & 446045 \\
$5\!f^2$   & 4 &  & 1.1198  & 69475  & 58035 & 461543 \\
$5\!f^2$   & 1 & C & 1.5000  & 75018  & 58132 & 449977 \\
$5\!f^2$   & 6 &  & 1.0296  & 78739  & 63175 & 460416 \\
$5\!f^2$   & 2 &  & 1.2672  & 89580  & 75041 & 465293 \\
$5\!f^2$   & 0 &  &         & 127521 & 114986 & 446376 \\
$5\!f\,6p$ & 3 &  & 0.9765  & 211414 &  212632 & 323435 \\
$6p^{2}$   & 1 &  & 1.4963  & 198879 &  213864 & -113277
\end{tabular}
\end{ruledtabular}
\end{table}

Our CI-only calculations showed the $6p^{2}\ (J=0)$ level to be the ground state, but adding MBPT corrections changes the level ordering such that $5\!f\,6p\ (J=3)$ is the ground state. Actually, Cf$^{16+}$ can be considered to have two ground states, since the decay from the metastable $6p^{2}\ (J=0)$ (G2 in \Tref{tab:Cf16+}) to the ground state (G1) has a lifetime greater than that of the nucleus itself. \Tref{tab:CfTrans} lists calculated matrix elements and strengths for some transitions of interest.

The two electron transitions between the G2 metastable state and the $5\!f^{2}$ states give maximal values of $q$: up to around $\sim 830\,000\,\cm$. Among these is the transition to $5\!f^{2}\ (J=1)$ (C in \Tref{tab:Cf16+}) with energy $\omega = 58\,132 - 5267 = 52\,865\,\cm$, which has a very high branching-ratio back to the G2 ``ground'' state. This level therefore potentially provides a method to ``recycle'' from G1 back to G2, although it should be noted that the G1$\,\rightarrow\,$C transition is rather weak. 

Another very interesting potential reference transition is the $\textrm{G2}\ (J=0)\,\rightarrow\,\textrm{B}\ (J=0)$ transition at $\omega = 46\,158\,\cm$, which is strongly forbidden but could be opened using Stark amplitude or hyperfine mixing of state B $(J=0)$ with C $(J=1)$. Such a transition would be very narrow and have strongly reduced systematic shifts, e.g. electric quadrupole, AC Stark, Zeeman shifts. It may, however, be too weak to excite by usual optical lasers.

All of the transitions discussed so far are positive shifters: the transition frequency increases with increasing $\alpha$. It is also possible to find negative shifters in Cf$^{16+}$, for example the transition between G1 and G2 is a strong negative shifter (assuming that the ordering of levels has been calculated correctly). However, this transition is extremely weak, and in practise may only occur via level mixing using a strong laser. A negative shifter which may be more useful is from the $5\!f^2\ (J=4)$ metastable state (A in \Tref{tab:Cf16+}; lifetime $\sim 10^{-1}$~s) via M1 transition to one of the $5\!f\,6p$ states above it. The larger of these has $q = -355\,000\,\cm$.

\section{Systematics and opportunities}
HCIs have some interesting features that are worth mentioning here. Firstly, electric dipole matrix elements are much smaller for HCIs than in neutral atoms since the E1 matrix element $\sim \langle r \rangle \sim \langle a_0/Z_\textrm{eff} \rangle$ where $a_0$ is the Bohr radius and $Z_\textrm{eff} \approx Z_\textrm{ion} + 1$ is the effective nuclear charge: the charge that the valence electron sees. Since the spacing between E1 levels in HCIs is larger by a factor $\sim Z_\textrm{eff}^2$, the static polarizability --- and hence black-body radiation shift --- of HCIs is reduced compared to near-neutral ions by a factor $\sim 1/Z_\textrm{eff}^4$.

The hyperfine structure in heavy HCIs is much larger than in neutral atoms, scaling as $\omega_\textit{hfs} \sim Z Z_\textrm{eff}^2$. The rate of M1 transitions within each hyperfine multiplet will scale as $\omega_\textit{hfs}^3$, which means that the lowest hyperfine state will be produced in reasonable time (order of a second). In californium, the hyperfine splitting of an $s$-wave or $p_{1/2}$-wave valence electron will be very sensitive to $\alpha$-variation because of the large $Z$. We define the fractional (relative) sensitivity $K_\textrm{rel}$ by $\delta\omega_\textit{hfs}/\omega_\textit{hfs} = K_\textrm{rel}\, \delta\alpha/\alpha$. Using formulas presented in Refs.~\cite{prestage95prl,flambaum06prc} we obtain $K_\textrm{rel} = 5.33$. Therefore, the hyperfine transitions form another positive shifting transition that can be used to place limits on $\alpha$-variation.

\begin{table*}[htb]
\caption{$q$ values, squared reduced transition matrix elements, $S$, and corresponding Einstein $A$-coefficients for transitions between selected states $i$ and $f$ in Cf$^{16+}$. The included $6p^{2}$ has longer lifetime than the nucleus.}
\label{tab:CfTrans}
\begin{ruledtabular}
\begin{tabular}{lcrlcrrcc}
\multicolumn{1}{c}{Config.} & $J_i$ & \multicolumn{1}{c}{$E_i\ (\cm)$} & 
\multicolumn{1}{c}{Config.} & $J_f$ & \multicolumn{1}{c}{$E_f\ (\cm)$} & 
\multicolumn{1}{c}{$q_{i\!f}\ (\cm)$} & 
\multicolumn{1}{c}{$S$} & \multicolumn{1}{c}{$g_f.A_{i\!f}~(\textrm{s}^{-1})$} \\
\hline
$5\!f\,6p$ & 3 & 0 & $6p^{2}$ & 0 & 5267 & -370928 & 0.92401$_{M3}$ & 4.3519\E{-18} \\
&&& $5\!f\,6p$ & 2 & 6104 & 106124 & 0.14553$_{M1}$ & 0.89281\\
&&& && & " & 0.35612$_{E2}$ & 0.00033798 \\
&&& $5\!f^{2}$ & 4 & 9711 & 414876 & 0.16895$_{M1}$ & 4.1731 \\
&&& && & " & 1.6768$_{E2}$ & 0.016216 \\
&&& $5\!f\,6p$ & 4 & 24481 & 162126 & 2.5836$_{M1}$ & 1022.5 \\
&&& && & " & 0.20237$_{E2}$ & 0.1993 \\
&&& $5\!f^{2}$& 2 & 24483 & 354444 & 0.0041938$_{M1}$ & 1.6601 \\%
&&& && & " & 0.079461$_{E2}$ & 0.078285 \\
&&& $5\!f\,6p$ & 3 & 25025 &  59395 & 0.071521$_{M1}$ & 30.235 \\%
&&& && &  " & 0.0022665$_{E2}$ & 0.0024914 \\
&&& $5\!f^{2}$ & 5 & 29588 & 451455 & 0.0011599$_{E2}$ & 0.0029459 \\
&&& $5\!f^{2}$ & 3 & 37467 & 393755 & 0.063566$_{M1}$ & 90.182 \\%
&&& && & " & 0.016359$_{E2}$ & 0.13528 \\
&&& $5\!f^{2}$ & 4 & 42122 & 319216 & 1.1631$_{M1}$ & 2344.7 \\
&&& && & " & 0.25404$_{E2}$ & 3.7727 \\
&&& $5\!f^{2}$ & 2 & 51471 & 446045 & 0.0082138$_{M1}$ & 30.211 \\%
&&& && & " & 0.016684$_{E2}$ & 0.6750 \\
&&& $5\!f^{2}$ & 4 & 58035 & 461543 & 0.13984$_{M1}$ & 737.27 \\%
&&& && & " & 0.028702$_{E2}$ & 2.1162 \\
&&& $5\!f^{2}$ & 1 & 58132 & 449977 & 0.0023199$_{E2}$ & 0.17249 \\
&&& $5\!f^{2}$ & 2 & 75041 & 465293 & 0.0036215$_{M1}$ & 41.278 \\
&&& && & " & 0.0022201$_{E2}$ & 0.59165 \\
$6p^{2}$ & 0 & 5267 & $5\!f\,6p$ & 2 & 6104 & 477052 & 0.54123$_{E2}$ & 2.4845\E{-8} \\
&&& $5\!f^{2}$ & 4 &  9711 & 785804 & 0.24444$_{E4}$ & 8.5359\E{-24} \\
&&& $5\!f^{2}$ & 2 & 24483 & 725372 & 0.019556$_{E2}$ & 0.0057377 \\
&&& $5\!f^{2}$ & 2 & 51471 & 816973 & 0.020833$_{E2}$ & 0.49127 \\
&&& $5\!f^{2}$ & 1 & 58132 & 820905 & 0.1195$_{M1}$ & 476.2 \\
&&& $5\!f^{2}$ & 2 & 75041 & 836221 & 0.00071052$_{E2}$ & 0.13159 \\
$5\!f\,6p$ & 2 & 6104 & $5\!f^{2}$ & 4 & 9711 & 308752 & 0.1839$_{E2}$ & 1.2569\E{-5} \\
&&& $5\!f\,6p$ & 4 & 24481 & 56002 & 0.028394$_{E2}$ & 0.0066653 \\
&&& $5\!f\,6p$ & 3 & 25025 & -46729 & 3.1428$_{M1}$ & 574.24 \\
&&& $5\!f\,6p$ & 3 & 25025 & " & 0.033026$_{E2}$ & 0.0089701 \\
&&& $5\!f^{2}$ & 1 & 58132 & 343853 & 0.0082881$_{M1}$ & 31.485 \\
&&& $5\!f^{2}$ & 1 & 58132 & " & 0.043495$_{E2}$ & 1.8571 \\
$5\!f^{2}$ & 4 & 9711 & $5\!f\,6p$ & 4 & 24481 & -252750 & 1.605$_{M1}$ & 139.5 \\
&&& && & " & 0.052382$_{E2}$ & 0.0041243 \\
&&& $5\!f^{2}$ & 2 & 24483 & -60432 & 1.5203$_{E2}$ & 0.11978 \\
&&& $5\!f\,6p$ & 3 & 25025 & -355481 & 0.096413$_{M1}$ & 9.3407 \\
&&& && & " & 0.0016877$_{E2}$ & 0.00015922 \\
&&& $5\!f^{2}$ & 5 & 29588 & 36579 & 8.3595$_{M1}$ & 1771.0 \\
&&& && " & 0.023016$_{E2}$ & 0.0079994 \\
&&& $5\!f^{2}$ & 3 & 37467 & -21121 & 0.06325$_{M1}$ & 36.484 \\
&&& && & " & 0.055057$_{E2}$ & 0.10159 \\
&&& $5\!f^{2}$ & 4 & 42122 & -95660 & 0.39244$_{M1}$ & 360.42 \\
&&& && & " & 0.26664$_{E2}$ & 1.0681 \\
&&& $5\!f^{2}$ & 6 & 44107 & 44471 & 0.00026242$_{E2}$ & 0.0014151 \\
&&& $5\!f^{2}$ & 2 & 51471 & 31169 & 0.035908$_{E2}$ & 0.51073 \\
&&& $5\!f^{2}$ & 4 & 58035 & 46667 & 0.0011045$_{M1}$ & 3.362 \\
&&& && & " & 0.0033787$_{E2}$ & 0.099715 \\
&&& $5\!f^{2}$ & 6 & 63175 & 45540 & 5.15\E{-6}$_{E2}$ & 0.00025212 \\
&&& $5\!f^{2}$ & 2 & 75041 & 50417 & 0.015122$_{E2}$ & 2.0155 \\
$5\!f\,6p$ & 4 & 24481 & $5\!f\,6p$ & 3 & 25025 & -102731 & 0.99502$_{M1}$ & 0.0043219 \\
$5\!f^{2}$ & 2 & 24483 & $5\!f^{2}$ & 1 & 58132 & 95533 & 0.024099$_{M1}$ & 24.766 \\
&&& && & " & 0.18074$_{E2}$ & 0.87324 \\
$5\!f\,6p$ & 3 & 25025 & $5\!f^{2}$ & 1 & 58132 & 390582 & 0.084098$_{E2}$ & 0.37462 \\
$5\!f^{2}$ & 3 & 37467 & $5\!f^{2}$ & 1 & 58132 & 56222 & 1.1076$_{E2}$ & 0.46748 \\
$5\!f^{2}$ & 0 & 51425 & $5\!f^{2}$ & 1 & 58132 & 68991 & 1.6793$_{M1}$ & 13.671 \\
$5\!f^{2}$ & 2 & 51471 & $5\!f^{2}$ & 1 & 58132 & 3932 & 1.0698$_{M1}$ & 8.5301 \\
&&& && & " & 0.58299$_{E2}$ & 0.00085652
\end{tabular}
\end{ruledtabular}
\end{table*}

\section{Conclusion}
In this \paper, we used the CI + MBPT method to calculate the energy levels for Cf$^{16+}$ and Cf$^{17+}$ highly charged ions. These ions were chosen because they lie at the $5\!f$ -- $6p_{1/2}$ crossing point on the thallium isoelectronic sequence, which allows for optical transitions between different configurations from the ground state. Our calculations have identified several transitions in Cf$^{16+}$ that have the largest $q$ values ever seen in such atomic systems, and include several positive shifters (with $q$ up to $\sim830\,000\,\cm$) and negative shifters (e.g. $5\!f^2\ (J=4)\,\rightarrow\, 5\!f6p\ (J=3)$ with $q=-355\,000\,\cm$). A comparison of clocks using these reference transitions would have a total sensitivity $\Delta q = q_+ - q_- \approx 1\,200\,000\,\cm$, around 23 times more sensitive than the Hg$^+$ clock and Al$^+$ clock comparison used to obtain the best current laboratory limit on $\alpha$-variation.

Trapping and cooling of HCIs remains a difficult experiment, however electron-beam ion trap technology continues to improve~\cite{draganic03prl,crespo08cjp,hobein11prl,beiersdorfer09pscr}, and we hope that the potential benefits of HCI clocks will continue to motivate further studies. Using californium certainly adds another layer of complexity to the experiment since it doesn't occur naturally and must be produced, for example, at accelerators such as GSI and LBNL.

\acknowledgments
This work was supported by the Australian Research Council. Supercomputer time was provided by an award under the Merit Allocation Scheme on the NCI National Facility at the Australian National University.

\appendix
\section{Transition matrix elements}
The CI+MBPT method produces wavefunctions $\left| I \right>$ which are linear combinations of Slater determinants of orbitals (we use atomic units $\hbar = e = m_e = 1$)
\begin{equation}
\psi_{n \kappa m} = \frac{1}{r}\begin{pmatrix} f_{n \kappa} \Omega_{\kappa m}
    \\ ig_{n \kappa}\Omega_{-\kappa m} \end{pmatrix} \,.
\end{equation}
To calculate the relativistic electric ($EJ$) and magnetic ($MJ$) multipole reduced matrix elements between two wavefunctions $\psi_i$ and $\psi_f$ we use the following formulae (see, e.g.~\cite{johnson07book}):
\begin{gather}
S_{MJ} = 4c^2 \left| \langle n_i \kappa_i || q_J^{(M)} || n_f \kappa_f \rangle \right|^2 \\
S_{EJ} = \left| \langle n_i \kappa_i || q_J^{(E)} || n_f \kappa_f \rangle \right|^2
\end{gather}
where
\begin{widetext}
\begin{gather}
\langle n_i\kappa_i || q_J^{(M)} || n_f\kappa_f \rangle
    = \frac{(2J + 1)!!}{k^J} \langle -\kappa_i || C^J || \kappa_f \rangle \int_0^\infty 
      \frac{\kappa_i + \kappa_f}{J + 1}\, j_J(kr) \left[ f_i(r) g_f(r) +  g_i(r) f_f(r) \right] dr \\
\begin{split}
\langle n_i\kappa_i || q_J^{(E)} || n_f\kappa_f \rangle 
  = \frac{(2J + 1)!!}{k^J} &\langle \kappa_i || C^J || \kappa_f \rangle \int_0^\infty 
     j_J(kr) \left[ f_i(r) f_f(r) + g_i(r) g_f(r) \right] + \\
    & j_{J+1}(kr) \left( \frac{\kappa_i - \kappa_f}{J + 1} \left[ f_i(r) g_f(r) + g_i(r) f_f(r) \right] +\left[ f_i(r) g_f(r) - g_i(r) f_f(r) \right]\right) dr \,.
\end{split}
\end{gather}
\end{widetext}
In these equations, 
\begin{eqnarray}
\langle \kappa_i || C^J || \kappa_f \rangle & = & (-1)^{j_i + 1/2} \sqrt{(2j_i + 1)(2j_f + 1)} \times\nonumber\\
&& \begin{pmatrix} j_i & j_f & J\\ -1/2 & 1/2 & 0 \end{pmatrix} %\times\nonumber\\
 \xi(l_i + l_f + J)
\end{eqnarray}
where
\begin{equation}
\xi(x) = \begin{cases}
1, & \text{for $x$ even}\\
0, & \text{for $x$ odd} \ ,
\end{cases}
\end{equation}
and $j_J(kr)$ is the spherical Bessel function of order $J$ with argument $kr = \frac{\left|\omega_i - \omega_f\right|}{c} r$. % the bracketed array is a Wigner-$3j$ matrix.
We also calculate Einstein $A$-coefficients using
\begin{equation}
A_J^{(\lambda)} = \frac{(2J + 2)(2J + 1) k^{2J + 1}}{J[(2J+1)!!]^2} \frac{\left|\langle I || Q_J^{(\lambda)} || F \rangle\right|^2}{(2J_F + 1)}
\end{equation}
where the capital letters replacing previous lower case letters indicate that the quantity should be summed over the entire many body system, for example
\begin{equation}
 Q_{JM}^{(\lambda)} = \sum_{if} (q_{JM}^{(\lambda)})_{i\!f}\, a_i^\dagger a_f \ .
\end{equation}

\bibliography{references}

\end{document}